\documentclass[a4paper, 11pt, fleqn]{article}
\usepackage{longtable}
\usepackage{graphicx}
\usepackage{amssymb,amsmath,epsfig,array}
\usepackage{xcolor}

\begin{document}

%\selectlanguage{russian}

\title{\bf{FG Sge: new multicolor photometry and short-term dust shell clearing in 2019 }}

\author{V.P.~Arkhipova$^1$, N.P.~Ikonnikova$^1$\footnote{e-mail: ikonnikova@sai.msu.ru}, V.I.~Shenavrin$^1$, M.A.~Burlak$^1$,
A.M.~Tatarnikov$^1$,
\\ D.Yu.~Tsvetkov$^1$, A.A.~Belinskii$^1$, N.N.~Pavlyuk$^1$, S.Yu.~Shugarov$^{1, 2}$}

\date{\it $^1${Lomonosov Moscow State University, Sternberg Astronomical Institute, Moscow, Russia\\
$^2$Astronomical Institute of the Slovak Academy of Sciences, Tatranska Lomnica, Slovakia}}

\renewcommand{\abstractname}{ }

\maketitle

\begin{abstract}

We present the results of a new stage of the long-term photometric study of FG~Sge which is a quickly evolving central star of the planetary nebula Hen~1-5. Our new observations carried out on the SAI MSU telescopes in the optical ($BVR_CI_C$) and infrared (IR) ($JHKLM$) regions in 2008--2021 and 2013--2021, respectively, allowed us to trace the evolution of the star's brightness in recent years. The most significant observations were performed in 2019 when the star suffered a short clearing of the dust shell and became visible in $BVR_C$. Based on the spectral energy distribution of FG~Sge in the 0.4--5~$\mu$m wavelength range we derived the dust shell parameters: the size of dust grains $a=0.01\mu$m, the inner radius temperature $T_{\text{dust}}=900$~K, optical depth $\tau (K)=0.5$ ($\tau (V)=4.5$), the total mass of dust $M_{\text{dust}}=7\cdot10^{-5} M_{\odot}$. After the short-term clearing of the dust shell in 2019, another dust structure was ejected that resulted in the star fading in all the observed bands. Based on the IR brightness and color curves, we estimated the dust depth growth in 2019--2020.

\bigskip


\end{abstract}

\section*{INTRODUCTION}

The unique central star of the planetary nebula He1-5, which is a fast evolving variable star FG~Sge, has been returning from the classic evolutionary track for a planetary nebula central star to the Asymptotic Giant Branch for more than 100 years. According to the state of the art, this was preceded by the last helium shell flash resulting in the star's rapid cooling and expansion with the bolometric luminosity staying quite constant. Long-year observations of the star led to the conclusion that FG~Sge went along the supergiant sequence from the B to K spectral type being a pulsating variable when crossing the instability strip and entered the R~CrB phase with powerful dust ejections in 1992, which is still lasting nowadays. Photometric observations of the star carried out since 1992 (Woodward et al. 1993; Arkhipova et al. 1994, 1996; Gonzalez et al. 1998; Tatarnikov and Yudin 1998; Tatarnikov et al. 1998; Arkhipova et al. 2003, 2009; Taranova and Shenavrin 2002, 2013; Rosenbush and Efimov 2015) confirm the formation of a dust circumstellar envelope as a result of dust condensation in the ejecta from the star.

In 2019 Fadeyev published his stellar evolution calculations for population I stars with masses on the main sequence $1M_{\odot}<M_{ZAMS}<1.5M_{\odot}$ up to the stage of a cooling white dwarf (Fadeyev, 2019). He showed that the final helium flash (LTP) occurs in post-AGB stars with initial masses $1.30\div1.32$ $M_{\odot}$. A simulated model was applied to describe the evolution of FG~Sge and the theoretical pulsation period was compared with the observational ones derived in different years. The maximum pulsation period $P = 117$~d determined for an evolved star with the initial mass $M_{ZAMS}=1.3M_{\odot}$ and the overshooting parameter $f=0.016$ was in a good agreement with an observational estimate $P = 115$~d (Arkhipova et al. 2003) derived later than 1992. Besides, Fadeyev determined the current mass $M=0.565 M_{\odot}$, temperature $T_{*}= 4445$~K and radius $R_{*}= 126 R_{\odot}$ of the star (Fadeyev, 2019).

In this work, we present the multicolor observations of FG~Sge carried out at various telescopes of SAI MSU in 2008-2021 in the 0.4-5~$\mu$m wavelength range and describe some recent episodes
in its history.

\section*{OBSERVATIONS}

\subsection*{$BVR_CI_C$-photometry in 2008-2021}

We carried out photometric observations of FG~Sge on several telescopes with various detectors in 2008-2021. In 2009-2015 we acquired some data with the 70-cm reflector AZT-2 based in Moscow and equipped with a Apogee Ap-7p CCD camera (M70). In 2008-2019, we also used the 60-cm reflector Zeiss-2 of the Crimean Astronomical Station of SAI MSU, equipped at different times with Apogee AP-47p (C60a), FLI PL 4022 (C60b) and Aspen CG42 (C60c) CCDs. A great amount of Johnson-Cousins $BVR_{C}I_{C}$ photometric data was obtained in 2019-2021 with the use of the new telescope RC600 of the Caucasus Mountain Observatory (CMO) of SAI MSU equipped with an Andor iKon-L  2048$\times$2048 CCD (the pixel size is 13.5~$\mu$m, 0.67~arcsec/pix)(for more detail, see Berdnikov et al. 2020). The Maxim DL software was used to perform observations as well as primary data reduction including dark correction, bias subtraction and flat fielding.

\begin{figure}[ht]
    \centering
    \includegraphics[scale=0.6]{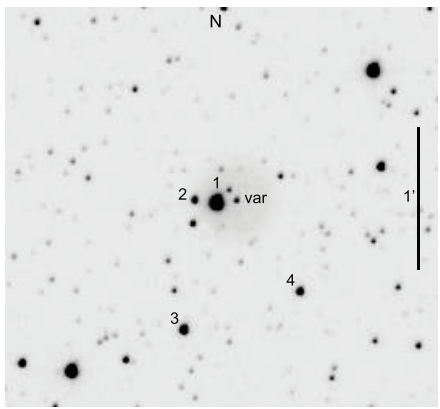}
    \caption{The vicinity of FG~Sge (var) with the comparison stars marked. The image was obtained on July 31, 2019 on the RC600 telescope in the $R_C$ band.}
    \label{fig1}
\end{figure}

{\sloppy FG~Sge varies significantly, so we used different comparison stars as photometric standards depending on the star's actual brightness. When FG~Sge was bright enough we usually used its optical companion (2MASS 20115664+2020031) located 7$''$ to the east. The companion's magnitudes were determined earlier in Arkhipova et al. (2003). The adopted magnitudes for the companion and fainter comparison stars are listed in Table~\ref{tabl1}. Fig.~\ref{fig1} shows the vicinity of FG~Sge with the comparison stars marked.

}

%---------------------fig.1-----------------------------------
%------------------------------------------------------------------

%---------------------tabl.1----------------------------------------
\begin{table}
\centering \caption{Comparison stars} \label{tabl1}
\begin{tabular}{|c|c|c|c|c|}

\hline
Designation&$B$&$V$&$R_C$&$I_C$\\
\hline
1&13.84&12.33&11.54&10.77\\
2&16.18&15.39&14.97&14.63\\
3&16.22&14.64&13.80&12.98\\
4&15.92&15.06&14.61&14.11\\

\hline
\end{tabular}
\end{table}
%------------------------------------------------------------------

The color terms needed to transform instrumental magnitudes to the standard $B$, $V$, $R_C$ and  $I_C$ ones were determined for some telescope-filter-detector combinations using the equations from Tsvetkov et al. (2006).

The RC600 data obtained in 2019--2021 were transformed to the standard Johnson-Cousins system according to the following expressions:

\begin{equation}
\begin{array}{c}

b = B - 0.061\cdot(B-V)\\
v = V + 0.027\cdot(B-V)\\
r = R_C+ 0.076\cdot(V-R_C)\\
i = I_C+ 0.074\cdot(R_C-I_C),\\

\end{array}
\end{equation}

where $bvri$ and $BVR_CI_C$ are the instrumental and standard system magnitudes, respectively. The color terms were calculated using observations of standards in M67 and the photometry of Chevalier and Ilovaisky (1991).

We performed aperture photometry using the WinFITS utility created by V.P.~Goranski. Our photometric precision is limited to $0.^{m}005$ for the bright state and is not worse than $0.^{m}05-0.^{m}10$ for deep minima.

In Table~\ref{tabl2} we present the photometry for FG~Sge obtained in 2008--2021 on the telescopes of SAI MSU.

\subsection*{$JHKLM$-photometry in 2013-2021}

{\sloppy $JHKLM$-photometry for FG~Sge has been performed on the 125-cm telescope of the Crimean Astronomical Station of SAI MSU with the InSb-photometer since 1985. A description of observational routine, instrumentation and observational data obtained in 1985--2008 can be found in Shenavrin et al. (2011). $JHKLM$-photometry for FG~Sge taken in 2009--2013 is presented in Taranova and Shenavrin (2013). Our new data acquired in 2013--2021 are given in Table~{\ref{tabl3}}.

}

\section*{PHOTOMETRY ANALYSIS}

\begin{figure}[ht]
    \centering
    \includegraphics[scale=1.2]{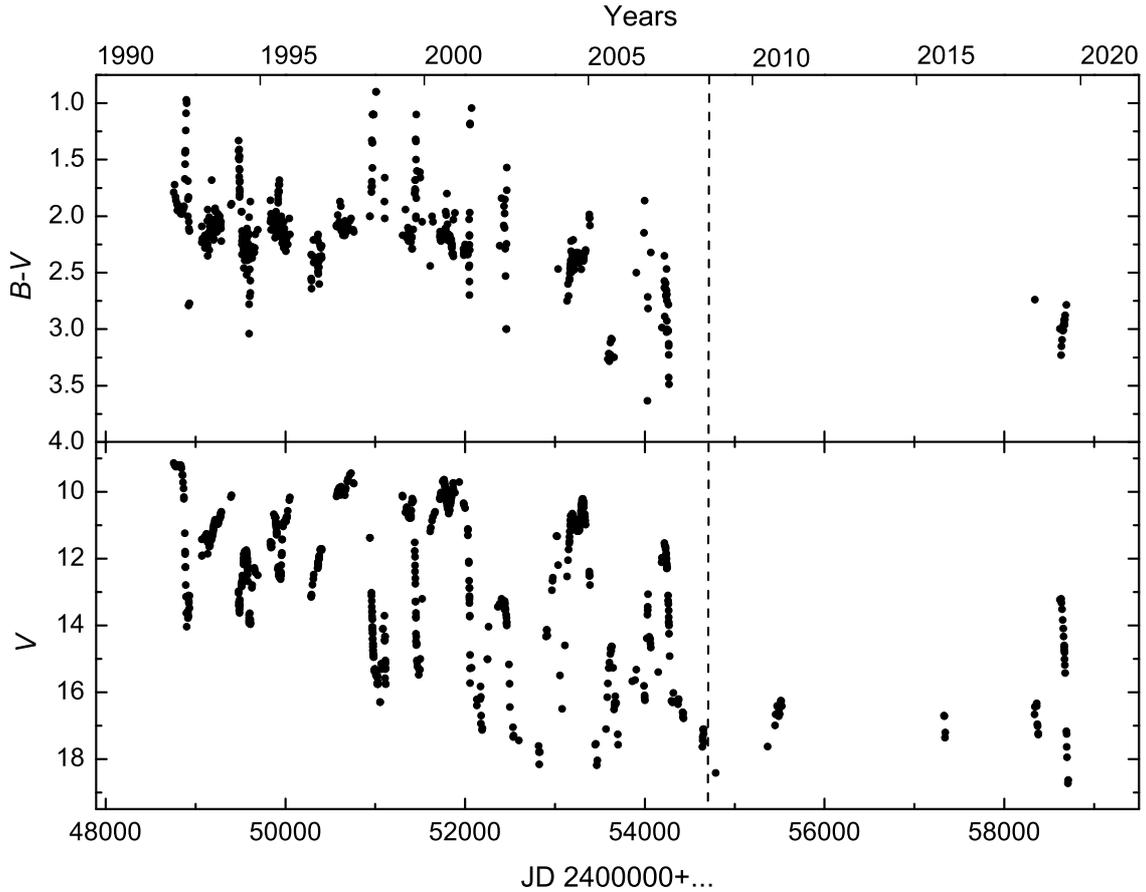}
    \caption{The $V$ light and $B-V$ color curves for 1992--2019. New data are located to the right from the vertical dashed line.}
    \label{fig2}
\end{figure}

Fig.~\ref{fig2} shows the $V$ light and $B-V$ color curves for 1992--2019 based on our photometry published in a series of papers in this journal in 1994--2009 and on new observations presented here.

%---------------------fig.2-----------------------------------
%------------------------------------------------------------------

In 1992--2009 the star displayed $B$ and $V$ light variations with an amplitude from 2 to 8 magnitudes, a general trend of decreasing mean brightness, a period of 115~d and a well-defined tendency to become bluer in the descending branch, which we discussed in detail in Arkhipova et al. (2009). But in 2010 the $V$ light fell to 16$^m$ and lower and we managed to measure the $B$ brightness only in 2019, as will be described below.

\begin{figure}[ht]
    \centering
    \includegraphics[scale=1.2]{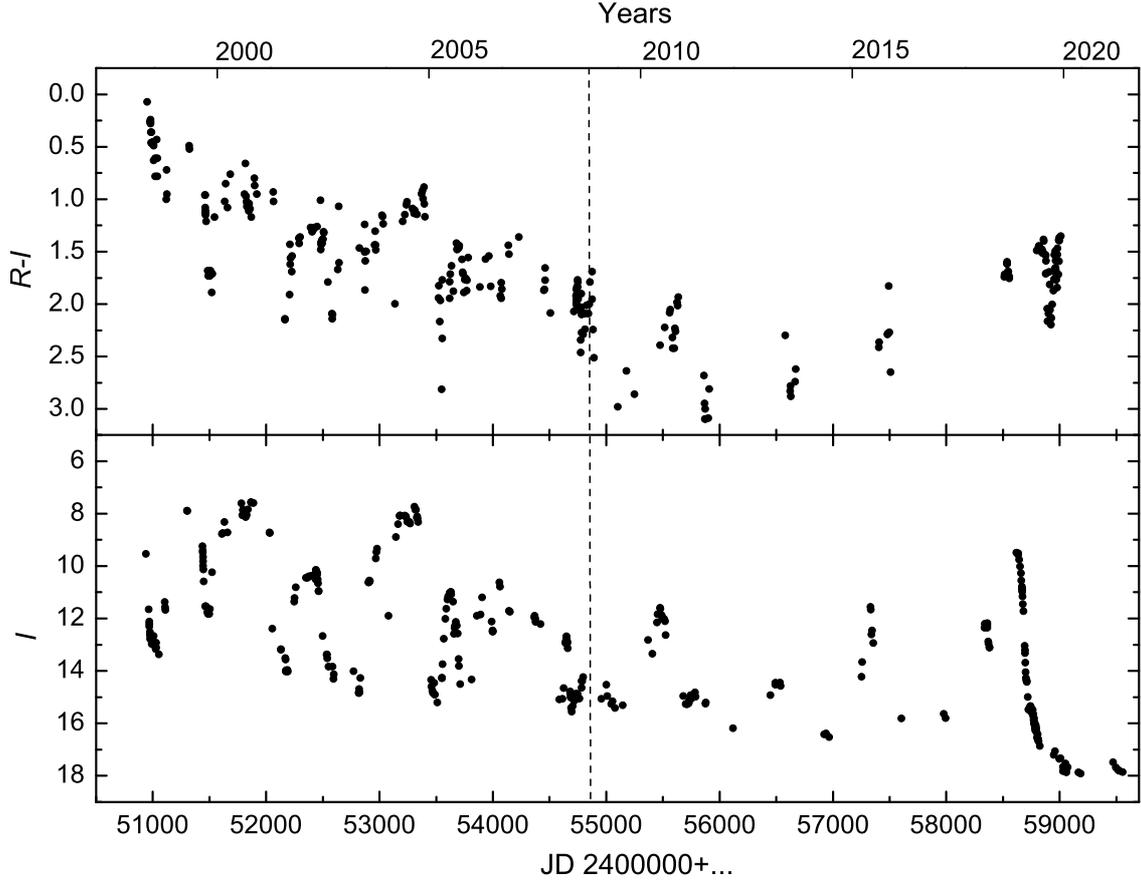}
    \caption{The $I$ light and $R-I$ color curves in 1998--2020. New data are located to the right from the vertical dashed line.}
    \label{fig3}
\end{figure}

Fig.~\ref{fig3} shows the $I$ light and $R-I$ color curves for 1998--2020 based on our data published in Arkhipova et al. (2003, 2009) and new observations taken in 2008--2020, the latter being transformed from the Cousins to Johnson system using the equations from Bessell (1979) and corrections obtained through simultaneous observations in the $R_C$ and $R$ bands.

%---------------------fig.3-----------------------------------
%------------------------------------------------------------------

Although FG~Sge was undetectable in $B$ and $V$, we managed to observe it in $R$ and $I$ all the time, and its $RI$ behavior before and after 2009 can be seen in Fig.~\ref{fig3} with a mean $I$ brightness fading up until 2019, a sudden rise to 10$^m$ in 2019 and dimming to 18$^m$ by early 2020. The $R-I$ color had been getting regularly redder from $0.^{m}5$ to $2.^{m}5$ by 2011, and had decreased to $1.^{m}5$ by the beginning of observations in 2019.

The pulsations were seen in $R$ and $I$ quite as well as in $V$ and $B$ but the $R-I$ color did not get bluer when the star was dimming.

\begin{figure}[ht]
    \centering
    \includegraphics[scale=1.5]{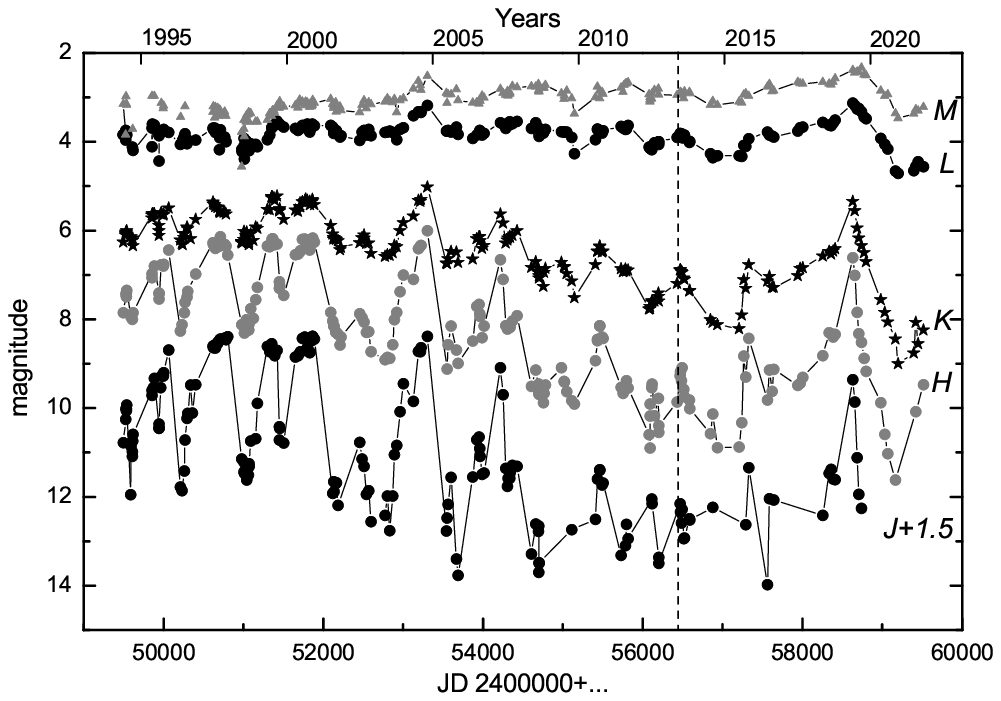}
    \caption{$JHKLM$ light curves for FG~Sge in 1994--2021. New data are located to the right from the vertical dashed line.}
    \label{fig4}
\end{figure}

\begin{figure}[ht]
    \centering
    \includegraphics[scale=1.5]{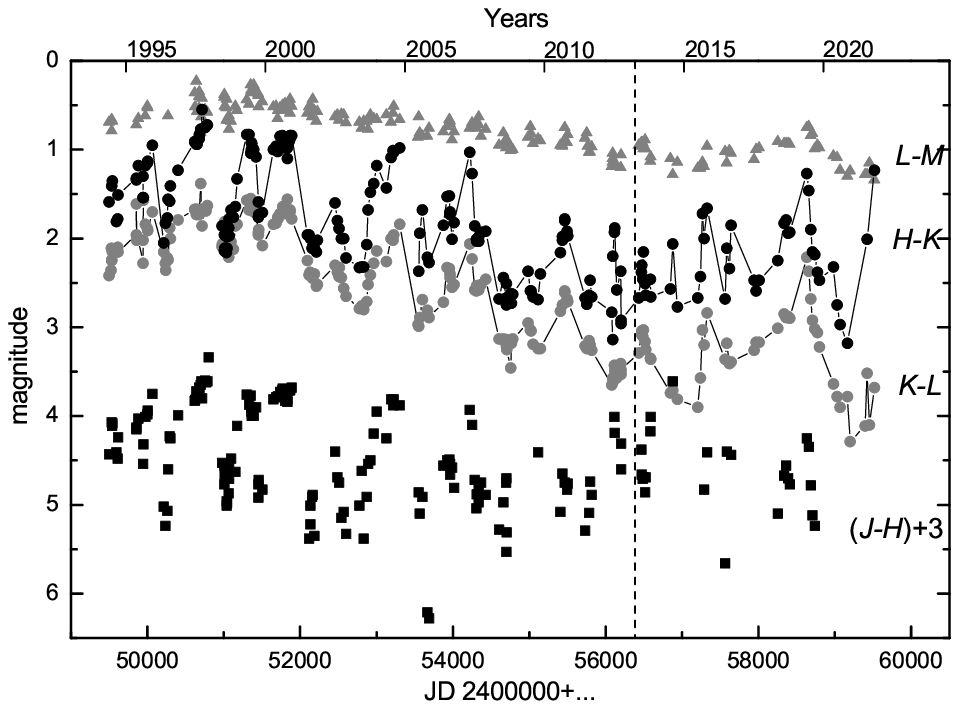}
    \caption{IR color curves for FG~Sge in 1994--2021. New data are located to the right from the vertical dashed line.}
    \label{fig5}
\end{figure}

Fig.~\ref{fig4} and \ref{fig5} show the IR light and color curves for FG~Sge in 1994--2021 based on observations reported in Shenavrin et al. (2011), Taranova and Shenavrin (2013) and our new data. A detailed description of IR data acquired for FG~Sge in 1993--2013 can be found in Taranova and Shenavrin (2013).

%---------------------fig.4-----------------------------------
%-------------------------------------------------------------

The object displayed large $J$ and $H$ variations of total light in 1994--2016 related to the pulsating variable star itself as well as to its dust shell; the variations show a general trend of dimming down to $J\sim13^{m}$ and $H\sim11^{m}$ and decreasing amplitude.

Then by 2019, the mean brightness in both bands suddenly increased by $3.^{m}5-4^{m}$. And in 2019 we observed a rapid drop in $H$ brightness of about 5 magnitudes and stopped seeing the object in $J$ that indicated a strong growth of the dust shell optical depth in both bands. In 2020 the object continued to decrease in $H$ to an all-time low brightness of $11.^{m}62$, whereas in 2021 it started brightening. The $K$ brightness evolution was similar to that of $H$.

The dust shell brightened in $L$ and $M$ by $1^{m}$ and $0.^{m}8$, respectively, from 1994 till 2019, and by the end of 2020 was fainter by $1.^{m}5-2.^{m}0$. The 2021 observations showed a slight brightening of the object in these bands.

%---------------------fig.5-----------------------------------

%-------------------------------------------------------------

It was shown by Taranova and Shenavrin (2013) and can be seen from Fig.~\ref{fig5} that the object was monotonously reddening in the available IR range 1.25--5~$\mu$m till 2013. The reddening in the $JHK$ bands was related to the growth of the dust shell optical depth in the line of sight, and in the farther IR region of 3.5--5~$\mu$m -- to the receding of dust structures from the heating star and the decreasing of the dust shell temperature, as a consequence.

The behavior of the $H-K$, $K-L$ and $L-M$ colors significantly changed after 2013. So, varing with an amplitude of up to 1 magnitude the $H-K$ color decreased in average from 2013 till early 2019. The object was also observed to get bluer in the 2.2--5~$\mu$m range ($KLM$). The $H-K$, $K-L$ and $L-M$ colors increased substantially during a rapid fall of brightness in all IR bands in 2019--2020. An inverse phenomenon occurred in 2021 -- the object was getting bluer when brighter.

\subsection*{Photometric observations of FG~Sge in 2019--2021}

We managed to detect FG~Sge in the optical range ($BVR_C$) in 2019 after several years of invisibility.

\begin{figure}[ht]
    \centering
    \includegraphics[scale=1.2]{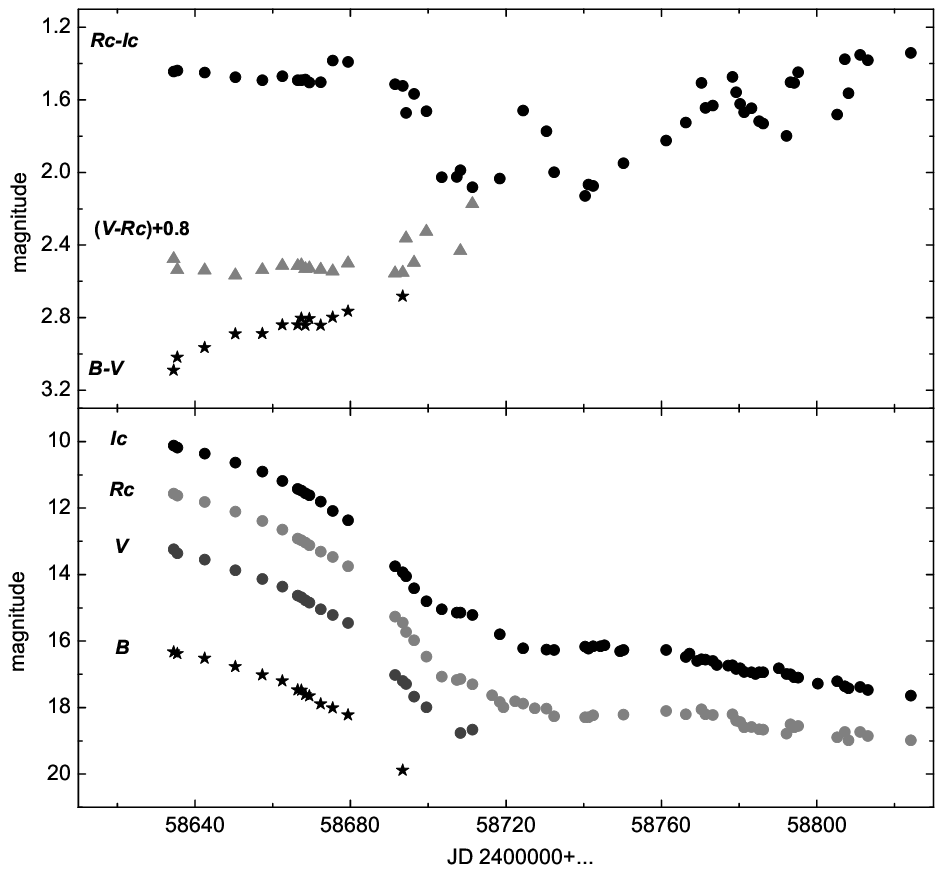}
    \caption{Light and color curves for FG~Sge based on the RC600 observations in 2019.}
    \label{fig6}

\end{figure}

The light and color curves for FG~Sge based on the RC600 observations in 2019 are rather well presented in Fig.~\ref{fig6}. The star was faint again in 2020--2021 and we could detect it only in the $I_C$ band at $18.^{m}1-18.^{m}5$.

%---------------------fig.6-----------------------------------

%-------------------------------------------------------------

In 2019, after having brightened to $16.^{m}2$ in $B$, the star declined by more than 3$^m$ in 75 days and became invisible for observations.

\begin{figure}[ht]
    \centering
    \includegraphics[scale=1.2]{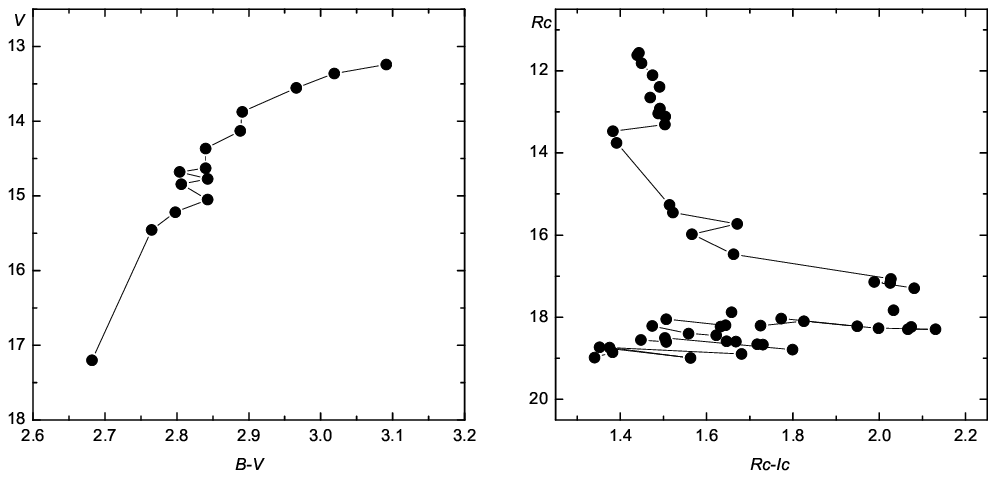}
    \caption{Color-brightness diagrams based on $BV$ and $R_CI_C$ observations.}
    \label{fig7}

\end{figure}

We managed to measure the $V$ brightness from $V\sim13^{m}$ to
$V\sim19^{m}$, whereas in the $R_C$ and $I_C$ bands we could trace the star declining from $11.^{m}56$ to $19^{m}$ and from $10.^{m}12$ to $18.^{m}$0, respectively. The fading was caused by the growth of extinction in the line of sight due to mass loss from FG~Sge via star wind and the formation of new dust structures. Fig.~\ref{fig7} shows the  $B-V$ and $R_C-I_C$ color versus $V$ and $R_C$ brightness correlations. $B-V$ got bluer by 0.$^m$4 and $V-R_C$ -- by 0.$^m$3 in 2019. The blueing is related to the presence of small particles in the newly formed dust shell that cause light scattering. When the brightness dropped to $18^{m}$, $R_C-I_C$ reddened from 1.$^m$4 to 2.$^m$0 and then recovered to $\sim1.^m4$.

%---------------------fig.7-----------------------------------

%-------------------------------------------------------------

We believe this behavior of $R_C-I_C$ to be due to an interaction of pulsations with the enhanced mass loss occurring in those days. It should be noted that back in 2001 the star displayed an unusual double deep minimum of about $V=17^{m}$, when the $V-I$ color grew from 2.$^m$5 to 3.$^m$5--3.$^m$7 and then fell to 2.$^m$5 twice (Arkhipova et al. 2003).

The observations of FG~Sge in 2019 revealed a rapid drop of brightness in optical range and a slower and smaller one in near IR (up to 5~$\mu$m).

In 2019 the IR brightness decreased by 3$^m$ in $J$, 4.$^m$5 in $H$ and 2.$^m$7 in $K$; the drop in $L$ and $M$ brightness of 1.$^m$0 and 0.$^m$8, respectively, was unusually large compared to the previous years. The star was fading in all the IR bands in 2020 and the $K$, $L$ and $M$ magnitudes reached the largest values ever observed by the end of 2020. The star turned red as never before. The $H-K$ and $L-M$ colors grew up to 3.$^m$2 and 1.$^m$3, respectively, on November 11, 2020 (JD2459165); $K-L$ reached a record value of 4.$^m$29 on December 15, 2020. A few observations carried out in 2021 showed a slight brightening in $KLM$ and a more significant one of 2$^m$ in $H$, whereas the $H-K$ and $K-L$ colors decreased and $L-M$ almost did not change.

\subsection*{Spectral energy distribution and the dust shell parameters estimation}

{\sloppy Multicolor photometry provides us an opportunity to derive absolute energy distribution
in the spectrum of FG~Sge and its dust shell if we manage to perform absolute calibration of all measured $BVR_CI_CJHKLM$ magnitudes (Strai\v{z}ys 1977; Koornneef 1983; Bessell 1998).

}

\begin{figure}[ht]
    \centering
    \includegraphics[scale=1.2]{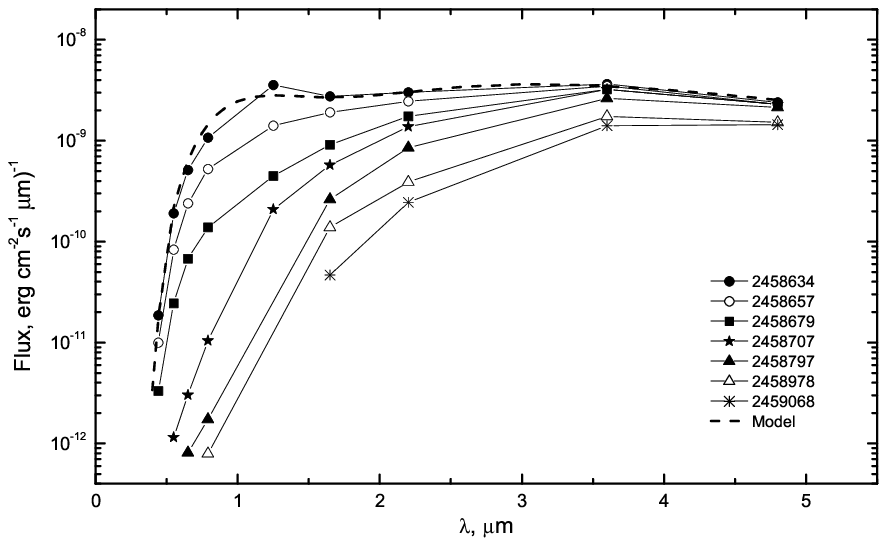}
    \caption{Spectral energy distribution of FG~Sge in 2019--2020 corrected for interstellar extinction. The JDs for the observations are given. A dashed line represents a model curve for JD2458634.}
    \label{fig8}

\end{figure}

In Fig.~\ref{fig8} we plot the absolute fluxes of the star in 2019--2020 dereddened with $E(B-V)=0.^{m}4$ (Arkhipova 1988).

%---------------------fig.8----------------------------------

%--------------------------------------------------------

One can see that the $V$, $R_C$ and $I_C$ fluxes decreased by almost 200 times in 73 days (JD2458634-2458707) and the $R_C$ and $I_C$ fluxes -- by 550 times in 163 days (JD2458634-2458797). The observed $V$, $R_C$, $I_C$ and, to a varying degree, $J$ and $H$ luminosity is a sum of the central star radiation attenuated by the dust shell plus the  scattered one. The $K$, $L$, $M$ fluxes are emitted by the dust shell.

The dust shell of FG~Sge has a complicated structure due to numerous episodes of mass loss and dust formation, and to the presence of clumps (which are responsible for the blueing of optical colors) that makes the modeling of the system a particularly challenging task. Nevertheless, one can estimate the basic parameters of the system with the simplest assumptions: a spherically symmetric shell was formed at a constant mass loss rate and contains small carbon grains (which are needed to provide a high degree of light scattering in the optical domain).

Based on the assumptions mentioned above, we simulated the spectral energy distribution for FG~Sge in the 0.4--5.0~$\mu$m wavelength range for JD2458634 using the CSDUST3 code (Egan 1988). For calculations we adopted the stellar parameters from Fadeyev (2019): $T_{*}=4445$~K and $R_{*}=126R_{\odot}$.

The resulting model curve is shown by a dashed line in Fig.~\ref{fig8}. The energy distribution observed before the abrupt decline in 2019 could be simulated only by assuming the presence of very small dust grains: $a=0.01\,\mu$m and less. The following parameters of the shell were derived: the inner radius $\sim40R_{*}$, the inner radius dust temperature 900~K, the optical depth $\tau(K)=0.5$ ($\tau (V)=4.5$), the total mass of dust $M_{\text{dust}}=7\cdot10^{-5} M_{\odot}$.

In 2019 FG~Sge displayed a significant decline in brightness (by more than 4 times) observed in a wide wavelength range reported here. This phenomenon together with the 5~$\mu$m flux decreasing could have been explained by a thick absorbing cloud having appeared in the line of sight; the cloud obscured the central star and a substantial part of the hot dust shell. But the large characteristic size of this part of the shell (tens of a.u.) and relatively short durance of the decline suggest large enough velocities for the cloud (of the order of 1000~km/s) which is implausible. Another scenario that can qualitatively explain the observed decline of brightness is that a new very dense spherically symmetric dust shell is formed around the central star. In order to avoid additional radiation in the observed wavelength range, the shell ought to be optically thick to its own near-IR emission. In this case, the already existing shell having lost the heating source will decrease its temperature and the maximum of its radiation will move to the mid-IR region.

Multicolor photometry allowed us to estimate the change in optical depth along the line of sight for the star and dust shell in 2019-2020. Based on the first observation of FG~Sge with the new forming shell in 2019 (JD2458634) and considering the interstellar extinction, we derived $J_0=7^{m}.54$, $H_0=6.^{m}41$ and $(J-H)_0=1.^{m}13$. Given that the star's normal brightness in $J$ is $J_0=6^{m}.59$ (Taranova and Shenavrin 2013), the total optical depth in $J$ has changed by nearly one magnitude.

After 108 days (JD2458738), our observations yielded $J_0=10^{m}.43$, $H_0=8.^{m}33$, $(J-H)_0=2.^{m}10$. This change in total light could well have been considered as a sufficient means for estimating the growth of the dust optical depth in $J$ in 108 days that would have been equal to $\Delta \tau(J)=A_{\lambda}/0.912=(10.43-7.54)/0.912=3.10$. But an optical light minimum (pulsational?) was observed around this date in the $V$, $R_C$, $I_C$ bands together with a clear blueing of the $B-V$ color; since the star could have its own light minimum, it would have been wrong to use the total $J$ magnitude for this date for optical depth calculations.

\begin{figure}[ht]
    \centering
    \includegraphics[scale=1.0]{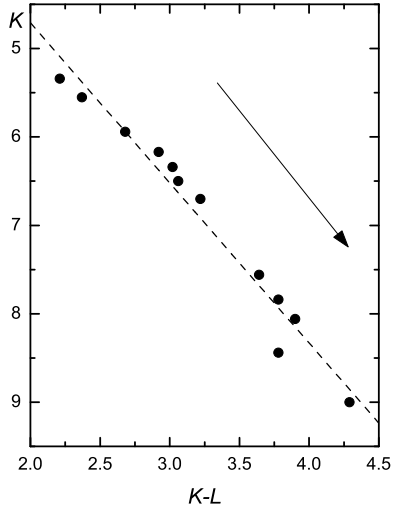}
    \caption{Color-brightness diagram for the 2019--2020 data. Black dots correspond to our observations. The dashed line represents linear fit. The reddening vector is shown by an arrow.}
    \label{fig9}

\end{figure}

The IR brightness had been decreasing until the end of 2020; as the object became invisible in $J$, we turned to the $KL$ bands to estimate the optical depth in 2020. The $(K-L)-K$ color-brightness diagram (Fig.~~\ref{fig9}) demonstrates the way the brightness decreased and the color increased in 2019--2020 (JD2458630-2459199). Linear fitting of data gives an expression: $\Delta K= (1.81\pm 0.36) \Delta (K-L)$. The correlation coefficient is 0.98. According to the normal extinction law for the IR region $\Delta K=2.0 \Delta (K-L)$ (Koornneef 1983), so, the slopes of the regression lines for the extinction law and for the observed relationship coincide with each other within the error limits. Based on this issue, we can conclude that the optical properties of grains in the new dust shell of FG~Sge are similar to those of the interstellar ones.

The $K$ brightness decreased by $3.^{m}66$ in 569 days (JD2458630-2459199) that gives an increase of the dust optical depth in $K$ of $\Delta \tau(K)=A_{\lambda}/0.912=3.66/0.912=4.0$.

%---------------------fig.9----------------------------------

%--------------------------------------------------------

\section*{CONCLUSIONS}

In this study, we carried out optical photometric observations of FG~Sge in 2008--2021 and near-IR ones in 2013--2021, confirmed the star to be active after the ejection of dust shell in 1992 and described the way the dust forms and evolves.

The star was fainter than $16^m$ in $V$ in 2010--2018 and invisible in $B$ till 2019, whereas it displayed two notable $R_C$ and $I_C$ light maxima. A short-time clearing of the dust shell occurred in the early 2019 that caused a rise in brightness to $V=13.^{m}2$ and $B=16.^{m}2$. Then, another dusty structure was ejected and the star's brightness dropped quite quickly in all the observed bands. The star stayed visible in $B$ for 75 nights, in $V$ for 93 nights and in $R_C$ for 206 nights and then again turned invisible in these bands for our telescopes. The use of the new RC600 telescope with a modern CCD camera since 2019 provided an increase in limiting brightness for the $I_C$ observations. So, we had been obtaining $I_C$ data during the deep light minimum till the end of 2021 and managed to measure the star's brightness of $18.^{m}1-18.^{m}5$ in 2020--2021.

It's worth mentioning that a clear blueing of the $B-V$ color occurred in 2019 during the descending branch of light curve. Such a phenomenon was observed during deep light minima more than once as reported, e.g., in Arkhipova et al. (1994, 2003, 2009). This behavior is also observed in R~CrB stars during deep minima and is a characteristic feature of young variable UX~Ori stars. Current studies interpret this phenomenon as a consequence of enhanced scattering of the central star radiation by the circumstellar dust. Grinin (1988) was the first to introduce a model explaining the blueing of two young irregular variables UX~Ori and WW~Vul during deep light minima; the detection of linear polarization for the UX~Ori stars confirmed the model. Based on the idea, Pugach (1988) proposed a phenomenological model for the deep light fading in R~CrB that described correctly not only the shape of color tracks in color-brightness diagrams but also the direction in which the star moved along the tracks.

Based on the spectral energy distribution obtained before FG~Sge started to descend into the 2019 deep minimum we derived the dust shell parameters for JD2458630: dust grain size $a=0.01\mu$m, the inner-radius temperature $T_{\text{dust}}=900$~K, the optical depth $\tau(K)=0.5$ ($\tau(V)=4.5$), the total mass of dust in the shell $M_{\text{dust}}=7\cdot10^{-5} M_{\odot}$.

The shell created in 2019 was characterized by an extraordinarily large optical depth. For the first time since FG~Sge had started to eject dust shells (i.e. since 1992) it displayed a significant light fading in mid-IR, too (the $L$ and $M$ bands). We have proposed a scenario which qualitatively explains the observed fading. In order to avoid additional radiation in the observed wavelength range, this new dust structure ought to be optically thick even to its own near-IR emission. In this case, the already existing shell having lost the heating source will decrease its temperature and the maximum of its radiation will move to the mid-IR region. Indeed the increase in optical depth caused the light fading in all the observed bands. From the first observation in 2019 till the end of 2020 (569 days) the optical depth in $K$ grew by $\Delta \tau(K)\sim4.0$.

In conclusion, we wish to draw attention to the new estimate of distance to FG~Sge obtained from the Gaia EDR3 results (Brown et al. 2021). Bailer-Jones et al. (2021) derived $D=2278_{-748}^{+1475}$ pc. This distance leads to the star's mean luminosity in the epoch preceding the start of dust ejections in 1992 of $\lg L/L_{\odot}=3.44$, if the mean value for FG~Sge in 1991 $V=9.^{m}2$ (Arkhipova et al. 2003) and $E(B-V)=0.^{m}4$ (Arkhipova 1988) are used. The star's $V$ brightness during the pulsational maximum of 1991 gives a luminosity higher by $0.^{m}3$: $\lg L/L_{\odot} = 3.74$. Our estimates of the FG~Sge luminosity obtained using the stellar parallax from EDR3 agree with the values assumed in Fadeyev (2019).

\bigskip

\bigskip

\section*{ACKNOWLEDGMENTS}

This study was carried out using the equipment purchased with the funds of M.V. Lomonosov Moscow State University Program of Development.
The work of one of the co-authors (SYS) was supported by grands from the Slovak Academy of Sciences VEGA 2/0030/21 and APVV-20-0148.

\section*{REFERENCES}

\begin{enumerate}

\item V.P. Arkhipova, Perem. Zvezdy {\bf22}, 631 (1988).

\item V.P. Arkhipova, G.V. Zaitseva, N.P. Ikonnikova, R.I. Noskova, and S.Yu. Shugarov, Astron. Lett. {\bf20}, 801 (1994).

\item V.P. Arkhipova, G.V. Zaitseva, N.P. Ikonnikova, R.I. Noskova, and S.Yu. Shugarov, Astron. Lett. {\bf22}, 750 (1996).

\item V.P. Arkhipova, N.P. Ikonnikova, G.V. Komissarova, R.I. Noskova, S.Yu. Shugarov, and V.F. Esipov, Astron. Lett. {\bf 29}, 763 (2003).

\item V.P. Arkhipova, V.F. Esipov, N.P. Ikonnikova, G.V. Komissarova, and S.Yu. Shugarov, Astron. Lett. {\bf 35}, 534 (2009).

\item C.A.L. Bailer-Jones, J.Rybizki, M. Fouesneau, et al., Astron. J. {\bf161}, 147 (2021).

\item L.N. Berdnikov, A.A. Belinskii, N.I. Shatskii, et al., Astron. Rep. {\bf 64}, 310 (2020).

\item M.S. Bessell, Publ. Astron. Soc. Pacific {\bf91}, 589 (1979).

\item M.S. Bessell, F. Castelli, and P. Planesas, Astron. Astrophys. {\bf 333}, 231 (1998).

\item A.G.A. Brown, A. Vallenari, T. Prusti, et al. (Gaia Collab.), Astron. Astrophys.  {\bf649}, A1 (2021).

\item C. Chevalier and S.A. Ilovaisky, Astron. and Astrophys. Suppl. Ser. {\bf 90}, 225 (1991).

\item M.P. Egan, C.M. Leung, and G.F. Spagna, Computer Physics Communication. {\bf48}, 271 (1988).

\item Yu.A. Fadeyev, Astron. Lett. {\bf 45}, 655 (2019).

\item G. Gonzalez, D.L. Lambert, G. Wallerstein, et al., Astrophys.J. Suppl. Ser. {\bf114}, 133 (1998).

\item V.P. Grinin, Soviet Astronomy Letters {\bf 14}, 27 (1988).

\item J. Koornneef, Astron.  Astrophys. {\bf 128}, 84 (1983).

\item A.F. Pugach, Soviet Astronomy {\bf 35}, 61 (1991).

\item A.E. Rosenbush and Yu.S. Efimov, Astrophysics, {\bf 58}, 46 (2015).

\item V.I. Shenavrin, O.G. Taranova, and A.E. Nadzhip, Astron. Rep. {\bf 55}, 31 (2011).

\item V. Strai\v{z}ys {\it Multicolor stellar photometry} (Vilnius: Mokslas Publishers, 1977)

\item O.G. Taranova and V.I. Shenavrin, Astron. Rep. {\bf 46}, 1010 (2002).

\item O.G. Taranova and V.I. Shenavrin, Astron. Lett. {\bf 39}, 781 (2013).

\item A.M. Tatarnikov and B.F. Yudin, Astron. Lett. {\bf 24}, 303 (1998).

\item A.M. Tatarnikov, V.I. Shenavrin, and B.F. Yudin, Astron. Rep. {\bf 42}, 377 (1998).

\item D.Yu. Tsvetkov, A.A. Volnova, A.P. Shulga, et al., Astron. Astrophys. {\bf 460}, 769 (2006).

\item C.E. Woodward, G.F. Lawrence, F.D. Gehrz, et al., Astrophys.J. {\bf 408}, 37, (1993).

\end{enumerate}

\begin{center}
\begin{longtable}{cccccc}
\caption{Photometric observations of FG~Sge in 2008--2021.}

\label{tabl2}\\

\hline
JD            &   $B$ &    $V$    &     $R_C$   &    $I_C$ & Telescope+\\
              &       &           &             &        & camera   \\
\hline

 \endfirsthead

\multicolumn{6}{c}
 {\tablename\ \thetable\ -- \textit{} } \\ \hline

JD            &   $B$ &    $V$ &  $R_C$   &    $I_C$ & Telescope+\\
              &       &        &          &          & camera   \\
   \hline

   \endhead

2454781.17    &   --  &   --  &   16.973  &   15.047  &   C60a    \\
2454787.18    &   --  &18.418 &   17.264  &   15.047  &   C60a    \\
2454795.15    &   --  &   --  &   17.394  &   14.907  &   C60a    \\
2454956.45    &   --  &   --  &   --      &   15.707  &   M70     \\
2455001.37    &   --  &   --  &   18.163  &   15.207  &   C60a    \\
2455009.42    &   --  &   --  &   --      &   15.607  &   C60a    \\
2455043.40    &   --  &   --  &   --      &   15.907  &   M70     \\
2455054.36    &   --  &   --  &   --      &   15.807  &   M70     \\
2455076.29    &   --  &   --  &   18.722  &   16.107  &   M70     \\
2455145.20    &   --  &   --  &   18.843  &   16.007  &   C60a    \\
2455367.52    &   --  &17.628 &   15.823  &   13.457  &   C60a    \\
2455407.51    &   --  &   --  &   16.183  &   13.987  &   C60a    \\
2455448.34    &   --  &16.995 &   14.832  &   12.777  &   M70     \\
2455454.37    &   --  &16.666 &   14.482  &   12.455  &   M70     \\
2455474.27    &   --  &16.405 &   14.502  &   12.207  &   M70     \\
2455477.29    &   --  &16.425 &   14.642  &   12.247  &   M70     \\
2455488.28    &   --  &16.715 &   14.882  &   12.487  &   M70     \\
2455494.26    &   --  &16.525 &   14.812  &   12.607  &   M70     \\
2455501.22    &   --  &16.655 &   14.842  &   12.607  &   M70     \\
2455514.21    &   --  &16.248 &   14.614  &   12.657  &   C60a    \\
2455518.21    &   --  &16.268 &   14.714  &   12.727  &   C60a    \\
2455525.20    &   --  &16.418 &   15.163  &   13.257  &   C60a    \\
2455678.51    &   --  &   --  &   --      &   15.607  &   M70     \\
2455701.46    &   --  &   --  &   --      &   15.907  &   M70     \\
2455712.45    &   --  &   --  &   --      &   15.927  &   M70     \\
2455725.52    &   --  &   --  &   --      &   15.887  &   C60a    \\
2455726.53    &   --  &   --  &   --      &   15.847  &   C60a    \\
2455728.49    &   --  &   --  &   --      &   15.867  &   C60a    \\
2455732.51    &   --  &   --  &   --      &   15.727  &   C60a    \\
2455744.36    &   --  &   --  &   18.264  &   15.607  &   C60a    \\
2455749.38    &   --  &   --  &   18.633  &   15.707  &   C60a    \\
2455753.43    &   --  &   --  &   18.823  &   15.747  &   C60a    \\
2455755.38    &   --  &   --  &   18.644  &   15.667  &   C60a    \\
2455757.39    &   --  &   --  &   18.733  &   15.657  &   C60a    \\
2455783.45    &   --  &   --  &   18.573  &   15.507  &   C60a    \\
2455789.41    &   --  &   --  &   18.453  &   15.667  &   C60a    \\
2455874.21    &   --  &   --  &   --      &   15.897  &   C60a    \\
2455879.20    &   --  &   --  &   --      &   15.857  &   C60a    \\
2456118.48    &   --  &   --  &   --      &   16.836  &   C60a    \\
2456447.44    &   --  &   --  &   17.872  &   15.599  &   M70     \\
2456490.39    &   --  &   --  &   17.944  &   15.137  &   C60a    \\
2456492.44    &   --  &   --  &   17.964  &   15.207  &   C60a    \\
2456493.47    &   --  &   --  &   17.994  &   15.137  &   C60a    \\
2456532.32    &   --  &   --  &   17.834  &   15.117  &   C60a    \\
2456537.37    &   --  &   --  &   17.843  &   15.247  &   C60a    \\
2456921.42    &   --  &   --  &   --      &   17.067  &   M70     \\
2456937.45    &   --  &   --  &   --      &   17.027  &   M70     \\
2456967.22    &   --  &   --  &   --      &   17.167  &   C60a    \\
2457253.43    &   --  &   --  &   17.273  &   14.886  &   C60b    \\
2457257.31    &   --  &   --  &   16.649  &   14.311  &   C60b    \\
2457330.12    &   --  &16.695 &   14.432  &   12.167  &   M70     \\
2457333.11    &   --  &16.715 &   14.532  &   12.277  &   M70     \\
2457340.28    &   --  &17.360 &   15.030  &   13.230  &   C60b    \\
2457344.14    &   --  &17.201 &   15.333  &   13.090  &   C60b    \\
2457355.50    &   --  &   --  &   16.202  &   13.577  &   M70     \\
2457602.50    &   --  &   --  &   --      &   16.460  &   C60b    \\
2457978.29    &   --  &   --  &   --      &   16.282  &   C60c    \\
2457992.27    &   --  &   --  &   --      &   16.447  &   C60c    \\
2458338.37    &   --  &16.665 &   14.687  &   12.974  &   C60c    \\
2458340.28    &19.169 &16.431 &   14.506  &   12.818  &   C60c    \\
2458360.43    &       &   --  &   14.358  &   12.789  &   C60c    \\
2458361.49    &   --  &16.335 &   14.459  &   12.871  &   C60c    \\
2458362.47    &   --  &16.417 &   14.556  &   12.970  &   C60c    \\
2458369.30    &   --  &16.956 &   15.161  &   13.502  &   C60c    \\
2458371.38    &   --  &17.012 &   15.246  &   13.584  &   C60c    \\
2458379.36    &   --  &17.234 &   15.451  &   13.741  &   C60c    \\
2458380.39    &   --  &17.267 &   15.456  &   13.730  &   C60c    \\
2458618.49    &16.228 &13.231 &   11.508  &   10.049  &   C60c    \\
2458634.50    &16.428 &13.198 &   11.498  &   10.072  &   RC600   \\
2458635.50    &16.471 &13.319 &   11.550  &   10.133  &   RC600   \\
2458642.50    &16.608 &13.514 &   11.742  &   10.315  &   RC600   \\
2458650.41    &16.851 &13.839 &   12.035  &   10.582  &   RC600   \\
2458657.38    &17.102 &14.093 &   12.321  &   10.849  &   RC600   \\
2458662.45    &17.286 &14.329 &   12.581  &   11.130  &   RC600   \\
2458666.42    &17.552 &14.595 &   12.845  &   11.370  &   RC600   \\
2458667.39    &17.563 &14.645 &   12.899  &   11.425  &   RC600   \\
2458668.40    &17.697 &14.737 &   12.969  &   11.500  &   RC600   \\
2458669.46    &17.729 &14.809 &   13.045  &   11.558  &   RC600   \\
2458672.35    &17.971 &15.011 &   13.240  &   11.755  &   RC600   \\
2458675.45    &18.095 &15.183 &   13.401  &   12.046  &   RC600   \\
2458679.41    &18.298 &15.422 &   13.686  &   12.318  &   RC600   \\
2458691.52    &   --  &   --  &   15.165  &   13.665  &   RC600   \\
2458693.42    &19.956 &17.171 &   15.350  &   13.842  &   RC600   \\
2458694.35    &   --  &17.251 &   15.641  &   13.959  &   RC600   \\
2458696.40    &   --  &17.635 &   15.882  &   14.322  &   RC600   \\
2458699.50    &   --  &17.95  &   16.380  &   14.704  &   RC600   \\
2458703.48    &   --  &   --  &   --      &   14.918  &   RC600   \\
2458707.38    &   --  &   --  &   --      &   15.018  &   RC600   \\
2458708.32    &   --  &18.729 &   17.044  &   15.026  &   RC600   \\
2458711.40    &   --  &18.629 &   17.224  &   15.086  &   RC600   \\
2458716.40    &   --  &   --  &   17.535  &   --      &   RC600   \\
2458718.43    &   --  &   --  &   17.738  &   15.672  &   RC600   \\
2458719.33    &   --  &   --  &   17.898  &   --      &   RC600   \\
2458722.40    &   --  &   --  &   17.718  &   --      &   RC600   \\
2458724.40    &   --  &   --  &   17.788  &   16.124  &   RC600   \\
2458727.40    &   --  &   --  &   17.932  &   --      &   RC600   \\
2458730.40    &   --  &   --  &   17.942  &   16.155  &   RC600   \\
2458732.35    &   --  &   --  &   18.173  &   16.144  &   RC600   \\
2458740.40    &   --  &   --  &   18.202  &   16.032  &   RC600   \\
2458741.27    &   --  &   --  &   18.198  &   16.096  &   RC600   \\
2458742.43    &   --  &   --  &   18.142  &   16.032  &   RC600   \\
2458744.35    &   --  &   --  &   --      &   16.068  &   RC600   \\
2458745.30    &   --  &   --  &   --      &   16.044  &   RC600   \\
2458749.37    &   --  &   --  &   --      &   16.218  &   RC600   \\
2458750.24    &   --  &   --  &   18.124  &   16.148  &   RC600   \\
2458761.22    &   --  &   --  &   18.006  &   16.163  &   RC600   \\
2458766.25    &   --  &   --  &   18.107  &   16.372  &   RC600   \\
2458767.22    &   --  &   --  &   --      &   16.291  &   RC600   \\
2458769.22    &   --  &   --  &   --      &   16.513  &   RC600   \\
2458770.33    &   --  &   --  &   17.958  &   16.458  &   RC600   \\
2458771.30    &   --  &   --  &   18.106  &   16.458  &   RC600   \\
2458773.22    &   --  &   --  &   18.136  &   16.500  &   RC600   \\
2458774.25    &   --  &   --  &   --      &   16.632  &   RC600   \\
2458777.17    &   --  &   --  &   --      &   16.650  &   RC600   \\
2458778.25    &   --  &   --  &   18.116  &   16.650  &   RC600   \\
2458779.20    &   --  &   --  &   18.304  &   16.748  &   RC600   \\
2458780.23    &   --  &   --  &   18.345  &   16.720  &   RC600   \\
2458781.27    &   --  &   --  &   18.502  &   16.828  &   RC600   \\
2458783.18    &   --  &   --  &   18.492  &   16.841  &   RC600   \\
2458784.18    &   --  &   --  &   --      &   16.905  &   RC600   \\
2458785.17    &   --  &   --  &   18.562  &   16.835  &   RC600   \\
2458786.18    &   --  &   --  &   18.573  &   16.832  &   RC600   \\
2458790.24    &   --  &   --  &   --      &   16.730  &   RC600   \\
2458792.21    &   --  &   --  &   18.694  &   16.879  &   RC600   \\
2458793.23    &   --  &   --  &   18.413  &   16.916  &   RC600   \\
2458794.19    &   --  &   --  &   18.506  &   17.006  &   RC600   \\
2458795.22    &   --  &   --  &   18.459  &   17.021  &   RC600   \\
2458800.21    &   --  &   --  &   --      &   17.188  &   RC600   \\
2458805.18    &   --  &   --  &   18.801  &   17.113  &   RC600   \\
2458807.18    &   --  &   --  &   18.643  &   17.283  &   RC600   \\
2458808.16    &   --  &   --  &   18.895  &   17.333  &   RC600   \\
2458811.17    &   --  &   --  &   18.640  &   17.305  &   RC600   \\
2458813.14    &   --  &   --  &   18.760  &   17.393  &   RC600   \\
2458824.17    &   --  &   --  &   18.890  &   17.568  &   RC600   \\
2458948.55    &   --  &   --  &   --      &   17.846  &   RC600   \\
2458958.51    &   --  &   --  &   --      &   17.710  &   RC600   \\
2458997.50    &   --  &   --  &   --      &   18.004  &   RC600   \\
2459006.50    &   --  &   --  &   --      &   17.976  &   RC600   \\
2459028.47    &   --  &   --  &   --      &   18.303  &   RC600   \\
2459030.50    &   --  &   --  &   --      &   18.469  &   RC600   \\
2459051.40    &   --  &   --  &   --      &   18.175  &   RC600   \\
2459055.50    &   --  &   --  &   --      &   18.446  &   RC600   \\
2459060.43    &   --  &   --  &   --      &   18.515  &   RC600   \\
2459070.43    &   --  &   --  &   --      &   18.318  &   RC600   \\
2459166.20    &   --  &   --  &   --      &   18.519  &   RC600   \\
2459185.16    &   --  &   --  &   --      &   18.560  &   RC600   \\
2459471.29    &   --  &   --  &   --      &   18.128  &   RC600   \\
2459514.20    &   --  &   --  &   --      &   18.424  &   RC600   \\
2459524.19    &   --  &   --  &   --      &   18.461  &   RC600   \\
2459555.13    &   --  &   --  &   --      &   18.501  &   RC600   \\

\hline

\end{longtable}
\end{center}

\begin{table}
\centering \caption{$JHKLM$-photometry of FG~Sge in 2013--2021.}
\label{tabl3}
\begin{tabular}{ccccccccccc}
\hline
JD&$J$&$\sigma J$&$H$&$\sigma H$&$K$&$\sigma K$&$L$&$\sigma L$&$M$&$\sigma M$\\
\hline

2456467.5&10.65&0.06&9.27&0.03&6.89&0.01&3.81&0.01&2.89&0.02\\
2456472.5&10.84&0.08&9.18&0.03&6.88&0.01&3.84&0.01&2.88&0.01\\
2456485.5&11.09&0.07&9.38&0.01&6.92&0.01&3.82&0.01&2.90&0.01\\
2456492.4&10.78&0.07&9.09&0.02&6.94&0.01&3.91&0.01&2.92&0.01\\
2456514.4&11.44&0.06&9.58&0.02&7.07&0.01&3.91&0.01&3.02&0.02\\
2456524.4&11.43&0.07&9.74&0.03&7.10&0.01&3.85&0.01&2.86&0.01\\
2456587.2&10.99&0.11&9.82&0.06&7.36&0.01&4.01&0.01&2.88&0.02\\
2456591.2&11.02&0.06&10.01&0.04&7.35&0.01&3.99&0.01&2.91&0.02\\
2456848.5&-&-&10.58&0.12&8.01&0.01&4.27&0.02&3.17&0.03\\
2456876.4&10.74&0.09&10.13&0.08&8.07&0.02&4.36&0.02&3.08&0.02\\
2456937.3&-&-&10.89&0.06&8.12&0.02&4.31&0.02&3.17&0.02\\
2457205.4&-&-&10.88&0.11&8.21&0.02&4.31&0.01&3.10&0.02\\
2457241.4&-&-&10.33&0.05&7.90&0.01&4.33&0.02&3.12&0.02\\
2457265.3&-&-&8.83&0.02&7.11&0.01&4.08&0.01&2.94&0.02\\
2457289.3&11.13&0.09&9.30&0.07&7.30&0.01&4.10&0.01&2.91&0.02\\
2457328.2&9.84&0.02&8.43&0.02&6.77&0.01&3.93&0.01&2.97&0.02\\
2457561.5&12.48&0.23&9.82&0.04&7.14&0.01&3.78&0.01&2.76&0.01\\
2457586.5&10.54&0.06&9.14&0.04&7.03&0.02&3.85&0.02&2.70&0.02\\
2457620.4&-&-&9.62&0.04&7.28&0.01&3.87&0.02&2.86&0.02\\
2457643.4&10.57&0.06&9.13&0.02&7.28&0.01&3.89&0.01&2.85&0.02\\
2457943.5&-&-&9.49&0.04&7.02&0.01&3.76&0.01&2.65&0.01\\
2457970.5&-&-&9.44&0.02&6.85&0.01&3.68&0.01&2.68&0.01\\
2458006.3&-&-&9.31&0.02&6.84&0.01&3.67&0.01&2.69&0.02\\
2458254.5&10.92&0.04&8.82&0.01&6.57&0.01&3.56&0.01&2.65&0.01\\
2458333.4&9.97&0.03&8.30&0.01&6.47&0.01&3.62&0.01&2.69&0.02\\
2458361.4&9.88&0.02&8.32&0.01&6.53&0.01&3.64&0.01&2.71&0.01\\
2458386.3&10.10&0.03&8.40&0.01&6.46&0.01&3.58&0.01&2.65&0.02\\
2458410.2&10.11&0.02&8.34&0.01&6.41&0.01&3.51&0.01&2.57&0.01\\
2458630.5&7.86&0.01&6.61&0.01&5.34&0.01&3.13&0.01&2.38&0.01\\
2458656.5&8.36&0.01&7.01&0.01&5.55&0.01&3.18&0.01&2.44&0.01\\
2458684.5&9.62&0.02&7.84&0.01&5.94&0.01&3.26&0.01&2.42&0.01\\
2458707.3&10.44&0.04&8.32&0.01&6.17&0.01&3.25&0.01&2.43&0.02\\
2458738.3&10.76&0.04&8.52&0.02&6.34&0.01&3.32&0.01&2.33&0.01\\
2458773.3&-&-&8.88&0.02&6.50&0.01&3.44&0.01&2.49&0.01\\
2458798.2&-&-&9.17&0.02&6.70&0.01&3.48&0.01&2.50&0.01\\
2458981.5&-&-&9.88&0.09&7.56&0.01&3.92&0.01&2.85&0.02\\
2459030.5&-&-&10.59&0.11&7.84&0.01&4.06&0.01&2.97&0.01\\
2459068.4&-&-&11.03&0.12&8.06&0.01&4.16&0.02&2.93&0.02\\
2459165.2&-&-&11.62&0.11&8.44&0.03&4.66&0.02&3.36&0.02\\
2459199.2&-&-&-&-&9.00&0.02&4.71&0.01&3.47&0.02\\
2459394.4&-&-&-&-&8.76&0.05&4.65&0.02&3.37&0.03\\
2459420.4&-&-&10.08&0.08&8.07&0.02&4.55&0.01&3.27&0.02\\
2459450.4&-&-&-&-&8.55&0.02&4.45&0.02&3.30&0.02\\
2459517.2&-&-&9.47&0.06&8.24&0.02&4.56&0.01&3.22&0.02\\

\hline
\end{tabular}

\end{table}

\end{document}